\documentclass[english,aps,superscriptaddress,preprint]{revtex4}
\usepackage[latin9]{inputenc}
\usepackage{float}
\usepackage{amsmath}
\usepackage{graphicx}
\usepackage{esint}
\usepackage[unicode=true]
{hyperref}

\makeatletter
\@ifundefined{textcolor}{}
{%
 \definecolor{BLACK}{gray}{0}
 \definecolor{WHITE}{gray}{1}
 \definecolor{RED}{rgb}{1,0,0}
 \definecolor{GREEN}{rgb}{0,1,0}
 \definecolor{BLUE}{rgb}{0,0,1}
 \definecolor{CYAN}{cmyk}{1,0,0,0}
 \definecolor{MAGENTA}{cmyk}{0,1,0,0}
 \definecolor{YELLOW}{cmyk}{0,0,1,0}
}

\makeatother

\usepackage{babel}
\begin{document}

\title{Subdiffraction-limited imaging based on longitudinal modes in a spatially
dispersive slab}

\author{Avner Yanai}

\affiliation{Department of Applied Physics, The Benin School of Engineering and
Computer Science, The Hebrew University of Jerusalem, Israel}

\author{Uriel Levy}

\email{ulevy@mail.huji.ac.il}

\affiliation{Department of Applied Physics, The Benin School of Engineering and
Computer Science, The Hebrew University of Jerusalem, Israel}
\begin{abstract}
It was proposed that a flat silver layer could be used to form a sub-diffraction
limited image when illuminated near its surface plasmon resonance
frequency {[}J. B. Pendry, Phys. Rev. Lett. 86, 3966 (2000){]}. In
this paper, we study the possibility of obtaining sub diffraction
resolution using a different mechanism, with no surface plasmons involved.
Instead, by taking into account the non-local response of a thin silver
slab, we show that longitudinal modes contribute to the formation
of a sub-diffraction limited image in a frequency regime above the
plasma frequency. The differences between these two distinct mechanisms
are studied and explained. 
\end{abstract}
\maketitle

\section{introduction}

In 2000, it was suggested that a thin planar metallic slab can perform
as a lens capable of producing an image with sub-diffraction limited
(SDL) resolution \cite{pendry}. In this seminal paper, Pendry showed
that by using a thin metallic slab, only few tens of nanometers thick,
which is excited with light near its surface plasmon (SP) resonance
frequency, an image with SDL size can be obtained, due to the high
momentum frequency components of the SP modes that are supported by
the interfaces between the metallic slab and the dielectric medium
surrounding it. This concept of a homogeneous slab performing as a
lens was a variant (perhaps more realistic) based on previous work
of Veselago that showed that ideal negative index slab, behaves as
a lens \cite{veselago}. Pendry also claimed that the resolution of
such an ideal negative index slab is not limited \cite{pendry}. It
is now common to coin the ideal negative index slab configuration
as a \textquotedblleft{}perfect lens\textquotedblright{}, while a
thin metallic layer is now defined as a \textquotedblleft{}poor man\textquoteright{}s
lens\textquotedblright{}. Both types of lenses became the subject
of debate and controversy \cite{collin}. The discussion was further
stimulated by the experimental demonstration of the \textquotedblleft{}poor
man\textquoteright{}s lens\textquotedblright{}, obtaining resolution
of the order of $\sim\lambda_{0}/10$ ($\lambda_{0}$ being the vacuum
wavelength) \cite{N_Fang,blaikie,Fang2}. Part of the debate regarding
the poor man\textquoteright{}s lens was concerned with the fact that
a momentum cutoff of the SP modes is inherent, due to the nonlocal
response of the metal permittivity function \cite{stockman,ramakrishna,ruppin-lens,ruppin_kempa}.
In the nonlocal description, the permittivity function is described
by $\varepsilon(\omega,\mathbf{k})$, i.e. it is a function of both
the optical frequency $\omega$ and the propagation constant $\mathbf{k}$,
due to temporal and spatial dispersion, respectively. For small $k$-vectors,
one can neglect spatial dispersion and assume that $\varepsilon(\omega,\mathbf{k})\approx\varepsilon(\omega)$.
However, for large $k$-vectors, the SP modes reach a momentum cutoff
resulting in a fundamental resolution limit for perfect imaging. A
detailed analysis, however, showed that the \textquotedblleft{}poor
man\textquoteright{}s lens\textquotedblright{} performance is very
similar under both local and nonlocal approximations, because the
ohmic losses inherent to the metal deteriorate the image even more
drastically as compared with nonlocal effects \cite{david_sci_rep}.
Nonlocal effects for metallic slabs are pronounced when the slab thickness
is less than the 10-nm regime. For these dimensions, it was shown
that the absorption spectrum above the plasma frequency $(\omega_{p})$
exhibits oscillations due to Fabry-Perot (FP) resonances of longitudinal
modes which are modes with zero magnetic field, and electric field
vector $\mathbf{E}$ parallel to the propagation vector $\mathbf{k}$.
Longitudinal modes satisfy the macroscopic Maxwell\textquoteright{}s
equations when $\varepsilon(\omega,\mathbf{k})=0$ \cite{agranovich_ginzburg,silin,forstmann-book}.
For several metals such as alkali metals and silver, a phenomenological
description of $\varepsilon(\omega,\mathbf{k})$ can be obtained by
employing the linearized hydrodynamic model \cite{feibelman,pitarke,boardman-ruppin,boardman-book,hanson,mortensen}.
This model successfully reproduces the appearance of longitudinal
mode resonances which were experimentally observed for thin layered
metals \cite{Anderegg} and offers a qualitative explanation for the
blue shifting of the localized SP resonance observed in silver nanoparticles
\cite{ciraci,raza2,pasha,toscano}. These results cannot be explained
with local models. However, the validity of this model is challenged
to account for quantum sized effects \cite{teperik-quantum-corrected,stella-hydro-quantum,esteban-quantum-corrected,monreal}.
In general, a full quantum-mechanical calculation is needed for an
accurate account of the non-local response. However, the longitudinal
mode resonance effect (described below) that is in the heart of this
paper, is fully accounted for by the hydrodynamic model. In this paper,
we propose a different physical concept for achieving SDL imaging,
taking advantage of the longitudinal modes within a thin metallic
slab that is described by the hydrodynamic model. The SDL imaging
occurs for discrete frequencies satisfying $\omega>\omega_{p}$. This
frequency regime is different from the original proposal of the \textquotedblleft{}poor
man\textquoteright{}s lens\textquotedblright{}, in which the incident
illumination frequency is in the vicinity of the SP resonance frequency
$\omega_{p}/\sqrt{2}$. Furthermore, the underlying physical mechanism
of the two approaches is different. In the original proposal, the
high-$k$ components of the source are reconstructed in the image
by the SP resonance surface modes which support high spatial frequency
components. In contrast, our approach is not based on surface modes.
Instead, the high-$k$ components are transferred to the image plane
by the longitudinal modes, which typically have propagation constants
more than an order of magnitude larger than the vacuum propagation
constant $k_{0}=2\pi/\lambda_{0}$. The paper is structured as follows.
In Sec. \ref{sec:Modal-analysis} we describe the modes inherent to
the nonlocal slab-lens geometry. In Sec. \ref{sec:Results-and-interpertation}
we present results based on this formalism. Sec. \ref{sec:conclusions}
concludes the paper.

\section{Modal analysis\label{sec:Modal-analysis}}

\begin{figure}
\begin{centering}
\includegraphics[scale=0.4]{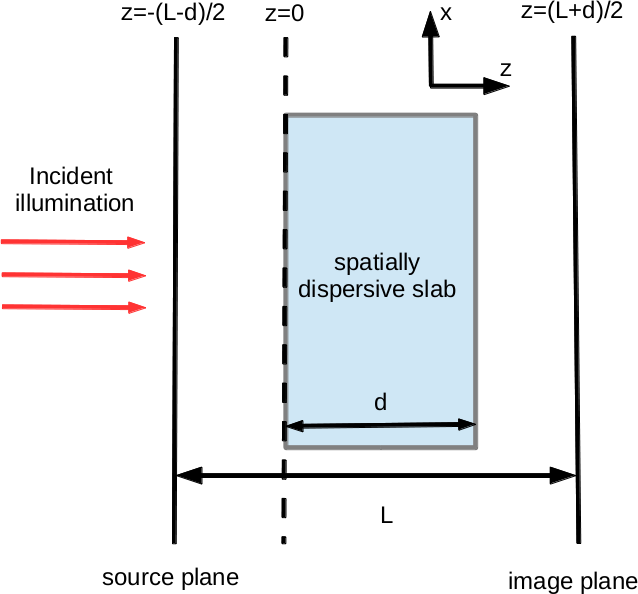}
\par\end{centering}

\caption{Schematics of a spatially dispersive slab of thickness $d$, surrounded
by vacuum.\label{fig:geometry}}
 
\end{figure}

Our lens geometry consists of a slab infinitely extending along the
$x$ and $y$ coordinates and bound between $z=0$ and $z=d$ (see
schematic in Fig. \ref{fig:geometry}). The image and source planes
are located at $z=-(L-d)/2$ and $z=(L+d)/2$ respectively. In our
analysis we assume that the slab is surrounded by vacuum. We consider
TM polarization illumination, i.e. $\mathbf{E}=E_{x}\hat{\mathbf{x}}+E_{z}\hat{\mathbf{z}}$,
for which plasmonic and longitudinal modes exist. The permittivity
of the slab is described by the hydrodynamic model. From this model
it follows that for transverse modes ($\mathbf{E}\perp\mathbf{k}$)
the permittivity function follows the well-known Drude model $\varepsilon_{T}(\omega)=1-\frac{\omega_{p}^{2}}{\omega^{2}+i\gamma\omega}$,
while for the longitudinal modes ($\mathbf{E}\parallel\mathbf{k}$)
the material response is given by $\varepsilon_{L}(\omega,\mathbf{k})=1-\frac{\omega_{p}^{2}}{\omega^{2}+i\gamma\omega-\beta^{2}k^{2}}$.
Here, $\gamma$ is the damping constant, and $\beta$ defines the
strength of the nonlocal response. The electric fields of the transverse
modes are given by $\mathbf{E_{T}}=E_{T}(-k_{z,T}/k_{x}\mathbf{\hat{x}+}\hat{\mathbf{z}})\exp(j[k_{x}x+k_{z,T}z])$,
and the propagation constant of the transverse mode satisfies the
dispersion relation $k_{z,T}^{2}=\varepsilon_{T}\omega^{2}/c^{2}-k_{x}^{2}$.
The electric fields of the longitudinal mode is given by\textbf{ }$\mathbf{E_{L}}=E_{L}(k_{x}/k_{z,L}\mathbf{\hat{x}+}\hat{\mathbf{z}})\exp(j[k_{x}x+k_{z,L}z])$,
and its propagation constant is given by $k_{z,L}^{2}=[\omega(\omega+i\gamma)-\omega_{p}^{2}]/\beta^{2}-k_{x}^{2}$,
which follows directly from the longitudinal mode dispersion relation
$\varepsilon_{L}(\omega,\mathbf{k})=0$. Because the spatially dispersive
slab supports twice the number of modes a non-spatially dispersive
slab does, an additional boundary condition (ABC) is needed other
than the two conventional conditions of continuity of the transverse
fields, in order to solve for all modal amplitudes \cite{agranovich_ginzburg,forstmann-book,Silveirinha,moreau}.
For an air/metal interface the set of boundary conditions we assume
are the continuity of $E_{x}$, $E_{z}$ and the normal current $J_{z}$.
Employing these boundary conditions and preserving the continuity
of $k_{x}$, one obtains the expression for the complex transmission
amplitude $T$ \cite{melnyk}. For completeness, the derivation for
the transmission amplitude is presented in the Appendix.

\section{Results and interpretation\label{sec:Results-and-interpertation}}

We now turn into studying the transmission through an Ag slab for
the following parameters \cite{blaber}: $\omega_{p}=9.6$ eV, $\gamma=(1/420)\omega_{p}$
, $\beta^{2}=(3/5)/v_{F}^{2}$ and $v_{F}=4.63\times10^{-3}c$. Here
$v_{F}$ and c are the Fermi velocity and the speed of light in vacuum
respectively. Possible surface roughness of the layer is not taken
into account in our simplified model (see Ref. \cite{feibelman} Sec.
4D). We continue our study assuming this set of parameters, although
the concepts derived below are general and support any spatially dispersive
material which can be described by the hydrodynamic model. In Fig.
2(a) we plot $\log(|T|^{2})$ as a function of $\omega$ and $k_{x}$
for the case of SP resonance, where the frequency $\omega$ is in
the vicinity of $\omega_{p}/\sqrt{2}$, assuming layer thickness of
$d=3$ nm. Two SP resonances with $k_{x}$ components below the light
line are observed. From previous studies \cite{khurgin,pendry2,collin,orenstein},
it is known that it is the interplay between the two modes that allows
for the reconstruction of an image with SDL resolution, due to the
very high transmission coefficients of plane waves with large $k_{x}$
(mind the logarithmic scale bar). 
\begin{figure}
\begin{centering}
\includegraphics[scale=0.1]{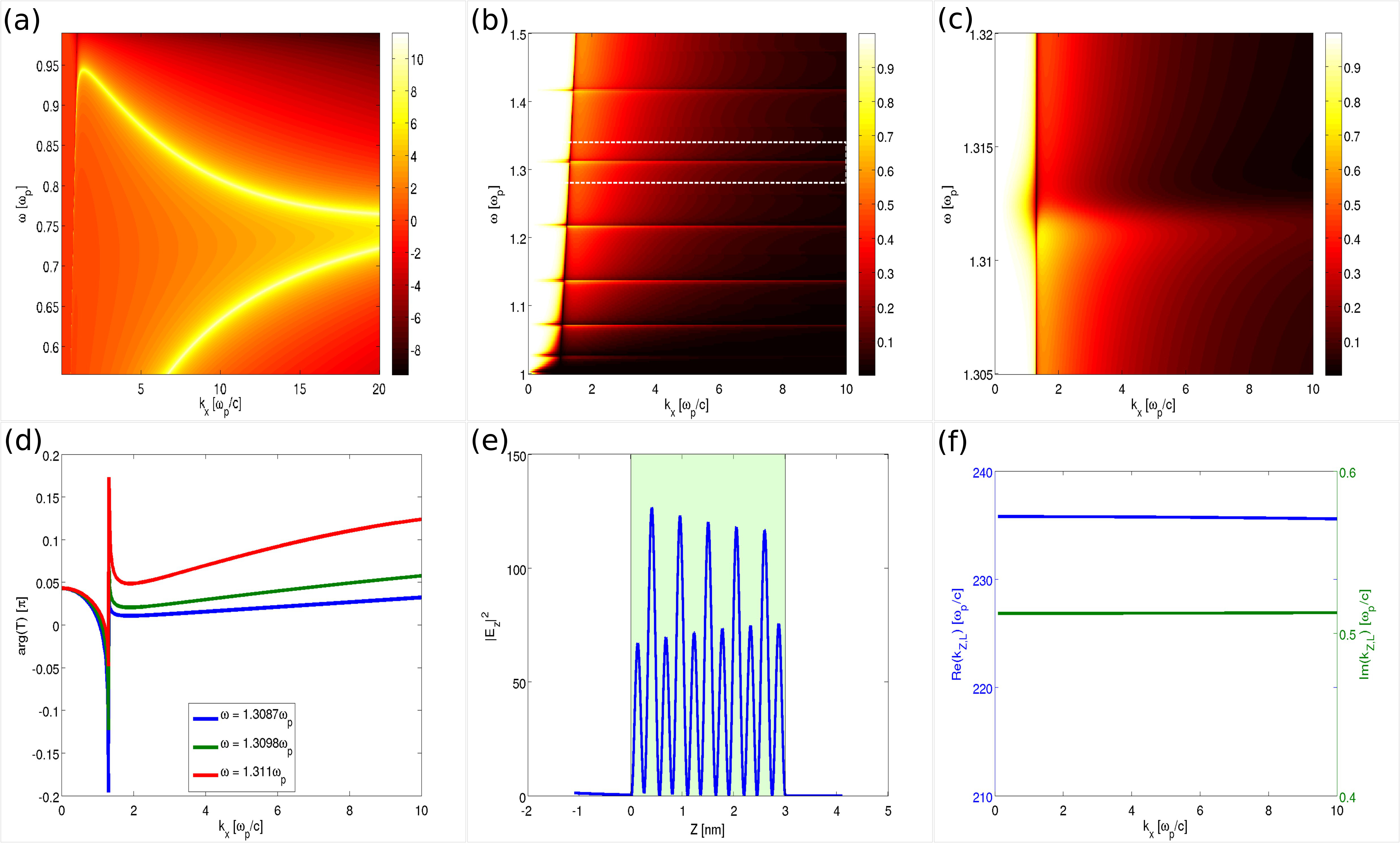}
\par\end{centering}

\caption{(a) $\log(|T|^{2}$) as function of $\omega$ and $k_{x}$ in the
vicinity of $\omega_{p}/\sqrt{2}$ (plasmonic resonance regime) and
$d=3$ nm (b) $|T|^{2}$ as function of $\omega$ and $k_{x}$ for
$\omega_{p}<\omega<1.5\omega_{p}$ (longitudinal mode FP regime) and
$d=3$ nm. (c) Detailed view of the FP resonance with $N=11$ near
$\omega=1.31\omega_{p}$ (d) The argument of the complex slab transmission
coefficient $T$ in units of $\pi$, calculated for several frequencies
near the $N=11$ FP resonance (e) Distribution of $|E_{z}|^{2}$ in
the slab (green background), calculated for $d=3$ nm, $\omega=1.31\omega_{p}$
and $k_{x}=6\omega_{p}/c$ (f) Real (blue line) and imaginary (green
line) parts of $k_{z,L}$ as function of $k_{x}$, calculated for
$\omega=1.31\omega_{p}$. \label{fig:comparision}}
 
\end{figure}
In Fig. \ref{fig:comparision}(b) we calculate $|T|^{2}$ assuming
the same slab thickness, however we now consider a different frequency
interval $(\omega_{p}<\omega<1.5\omega_{p}$). In this frequency range,
$k_{z,L}$ becomes predominantly real. Moreover, for the assumed ultrathin
slab dimensions, the longitudinal modes exhibit distinguishable FP
resonances at discrete frequencies satisfying $k_{z,L}d/\pi=N$, where
$N$ is an odd number (Ref. \cite[ section 3.2]{forstmann-book}).
A detailed view of one of these FP modes ($N=11$) at $\sim1.31\omega_{p}$,
is shown in Fig. \ref{fig:comparision}(c). At these resonance frequencies,
the spectrum of $|T|^{2}$ shows transmission peaks, which are nearly
uniform in frequency over a large range of $k_{x}$ values. Therefore,
the transmission spectrum at one of these resonant frequencies seems
to be more appealing for reconstruction of an SDL image (which typically
contains a large span of $k_{x}$ values) as compared to the SP resonance
case, which is more dispersive in nature, as shown in Fig. \ref{fig:comparision}(a).
On the other hand, the overall transmission magnitude of these longitudinal
modes is lower as compared to the SP modes (note that Fig. \ref{fig:comparision}(a)
is in logarithmic scale while Fig. \ref{fig:comparision}(b) is not!).
In addition to the FP transmission peaks which are evident for large
values of $k_{x}$, we also notice in Fig. \ref{fig:comparision}(b)
the existence of a region with nearly uniform transmission for low
values of $k_{x}$ which are above the light cone. In this region
the transmission is $\sim1$, besides discrete dips in transmission
corresponding to the FP modes. The FP modes exist mathematically for
$\omega>\omega_{p}$, but since $\text{Im}(k_{z,L})$ grows with $\omega$,
the resonance effect diminishes with the frequency. We next turn into
a further comparison between the two approaches, with the goal of
identifying the inherent properties and advantages of each approach. 

To perfectly image a line source, one needs both the transmission
magnitude and the transmission phase to be constant with $k_{x}$
(Ref. \cite{chew} Sec. 2.2.1). In Fig. \ref{fig:comparision}(d)
we plot the normalized phase of transmission $\arg(T)/\pi$ as a function
of $k_{x}$ for several frequencies near $1.31\omega_{p}$, to estimate
how uniform is the phase for different spatial frequencies. In Fig.\ref{fig:comparision}(e)
we plot the magnitude of the $E_{z}$ in the slab for a specific FP
resonance of the longitudinal modes ($\omega=1.31\omega_{p}$ and
$k_{x}=6\omega_{p}/c$). In Fig. \ref{fig:comparision}(f) we plot
the dependence of the real and imaginary part of $k_{z,L}$, as a
function of $k_{x}$, for $\omega=1.31\omega_{p}$. It is seen that
the variation in $k_{z,L}$ is negligible ($\sim0.1$\% in its real
value and$\sim1$\% in its imaginary value). This is primarily due
to its large value, which is the key reason for the FP transmission
resonances shown in Fig. 2(b) being nearly flat with $k_{x}$. 

To estimate the imaging capabilities of the lens, we define two point
spread functions (PSF), by calculating the transmission function at
the image plane obtained by summation of an infinite number of plane
waves according to: 
\begin{multline}
PSF_{E_{x}}(x,\omega)=\int_{0}^{\infty}-(k_{z0}/k_{x})T(k_{x},\omega)\exp(j[L-d]k_{z0}+jk_{x}x)dk_{x}+\\
\int_{0}^{\infty}-(k_{z0}/k_{x})T(k_{x},\omega)\exp(j[L-d]k_{z0}-jk_{x}x)dk_{x}\label{eq:PSFEx}
\end{multline}

\begin{multline}
PSF_{E_{z}}(x,\omega)=\int_{0}^{\infty}T(k_{x},\omega)\exp(j[L-d]k_{z0}+jk_{x}x)dk_{x}-\\
\int_{0}^{\infty}T(k_{x},\omega)\exp(j[L-d]k_{z0}-jk_{x}x)dk_{x}\label{eq:PSFEz}
\end{multline}
Here $PSF_{E_{x}}$and $PSF_{E_{z}}$ correspond to the PSF of the
$E_{x}$ and $E_{z}$ electric field components respectively. The
vacuum propagation constant $k_{z0}$ is defined by $k_{z0}^{2}=\omega^{2}/c^{2}-k_{x}^{2}.$
Equations \eqref{eq:PSFEx} and \eqref{eq:PSFEz} are defined for
the case of $E_{x}$ polarized illumination. Following these definitions,
we also define the total PSF in the image plane as: 

\begin{equation}
PSF_{T}=|PSF_{E_{x}}|^{2}+|PSF_{E_{z}}|^{2}\label{eq:PSFT}
\end{equation}
\begin{figure}
\begin{centering}
\includegraphics[scale=0.08]{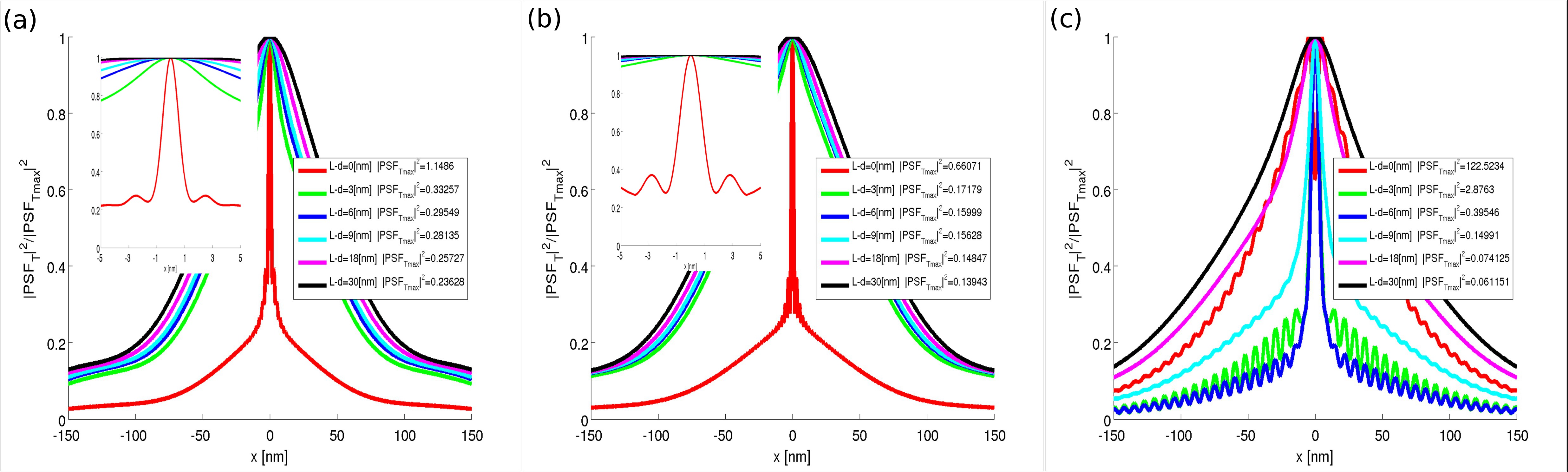}
\par\end{centering}

\caption{Normalized $PSF_{T}(x)$ calculated by Eq. \eqref{eq:PSFT}, as a
function of the transverse coordinate for several imaging distances,
assuming slab thickness of $d=3$ nm . The imaging distance $L-d$
and the normalization factor are shown in the legend. (a) $\omega=1.31\omega_{p}$.
(b) $\omega=1.0681\omega_{p}$. The insets show a magnification of
the PSF in the center. (c) $\omega=0.7071\omega_{p}$.\label{fig:PSF_Ag_3}}
 
\end{figure}
\begin{figure}
\begin{centering}
\includegraphics[scale=0.08]{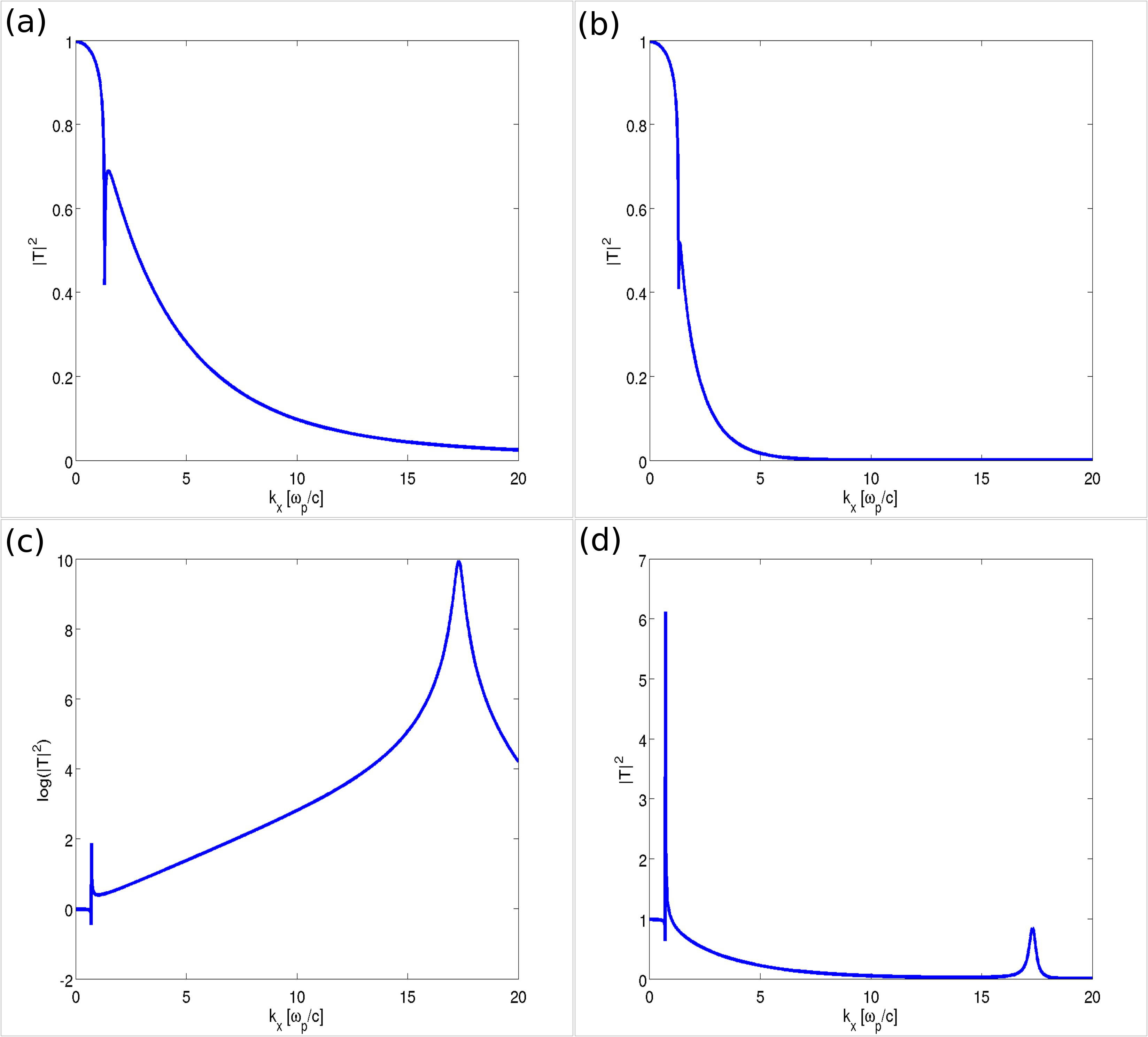}
\par\end{centering}

\caption{Transmission magnitude as a function of $k_{x}$. (a) $\omega=1.31\omega_{p}$
and $L-d=0$ (b) $\omega=1.31\omega_{p}$ and$L-d=6$ nm (c) $\omega=0.7071\omega_{p}$
and $L-d=0$ (d) $\omega=0.7071\omega_{p}$ and $L-d=6$ nm.\label{fig:T_vs_kx}}
 
\end{figure}
In Fig. \ref{fig:PSF_Ag_3} the normalized transmission magnitude
$|PSF_{T}|^{2}/|\text{max}(PSF_{T})|^{2}$ in the image plane is plotted
for several imaging distances $(L-d)$. The legend shows the normalization
factor $|\text{max}(PSF_{T})|^{2}$ for each imaging distance. Fig.
\ref{fig:PSF_Ag_3}(a,b) show the results obtained for the longitudinal
mode FP resonance with $N=11$ ($\omega=1.31\omega_{p}$) and $N=5$
($\omega=1.0681\omega_{p}$), respectively. Fig. \ref{fig:PSF_Ag_3}(c)
is calculated for $\omega=0.7071\omega_{p}$, corresponding to the
plasmonic resonance. For the case where the source is positioned at
the surface of the slab ($L-d=0$, red line) the longitudinal mode
FP mechanism results in an ultra-narrow FWHM of less than 2 nm, for
both FP resonances. However, when the imaging distance increases,
the FWHM increases and substantial broadening of the $PSF_{T}$ is
observed. We note that the PSFs of the two FP resonances (Figs. \ref{fig:PSF_Ag_3}(a,b))
are relatively similar. Once again, this is the outcome of the relatively
flat dispersion curves of the FP modes. In contrast, the plasmonic
resonance (Fig. \ref{fig:PSF_Ag_3}(c)) behaves very differently.
One may observe that the $PSF_{T}$ is narrower when the source is
slightly away of the surface, ($L-d=3$ and $6$ nm, green and blue
lines, respectively which have PSF FWHM of $\sim$6 nm) compared with
the case where the source is located on the surface ($L-d=0$). The
question arises, why for the plasmonic resonance case, the PSF narrows
with increasing imaging distance, while for the longitudinal mode
case, the PSF always broadens with the increase in imaging distance.
The underlying reason is revealed from the transmission spectrum,
calculated for several imaging distances. In Fig. \ref{fig:T_vs_kx}(a,b),
we plot the transmission magnitude as a function of $k_{x}$ of the
$N=11$ FP resonance at $\omega=1.31\omega_{p}$ for $L-d=0$ and
$L-d=6$ nm. It is seen that for $k_{x}$ components larger than the
vacuum light line (i.e. $k_{x}>1.31\omega_{p}/c$) we obtain a rapid
decay of the transmission for a distance of $6$ nm, compared with
the $L-d=0$ case. This observation explains the broader PSF for the
latter case (Fig. \ref{fig:PSF_Ag_3}(a), blue vs. red curve). 

The transmission spectrum is also useful for explaining the PSF results
of the plasmonic resonance. In Fig. \ref{fig:T_vs_kx}(c,d), it is
seen that for $L-d=6$ nm the transmission spectrum is more flat as
compared with the $L-d=0$ case (mind that Fig. \ref{fig:T_vs_kx}(c)
is in logarithmic scale). Therefore, the PSF becomes narrower (see
Fig. \ref{fig:PSF_Ag_3}(c), blue vs. red line). 

\begin{figure}
\begin{centering}
\includegraphics[scale=0.08]{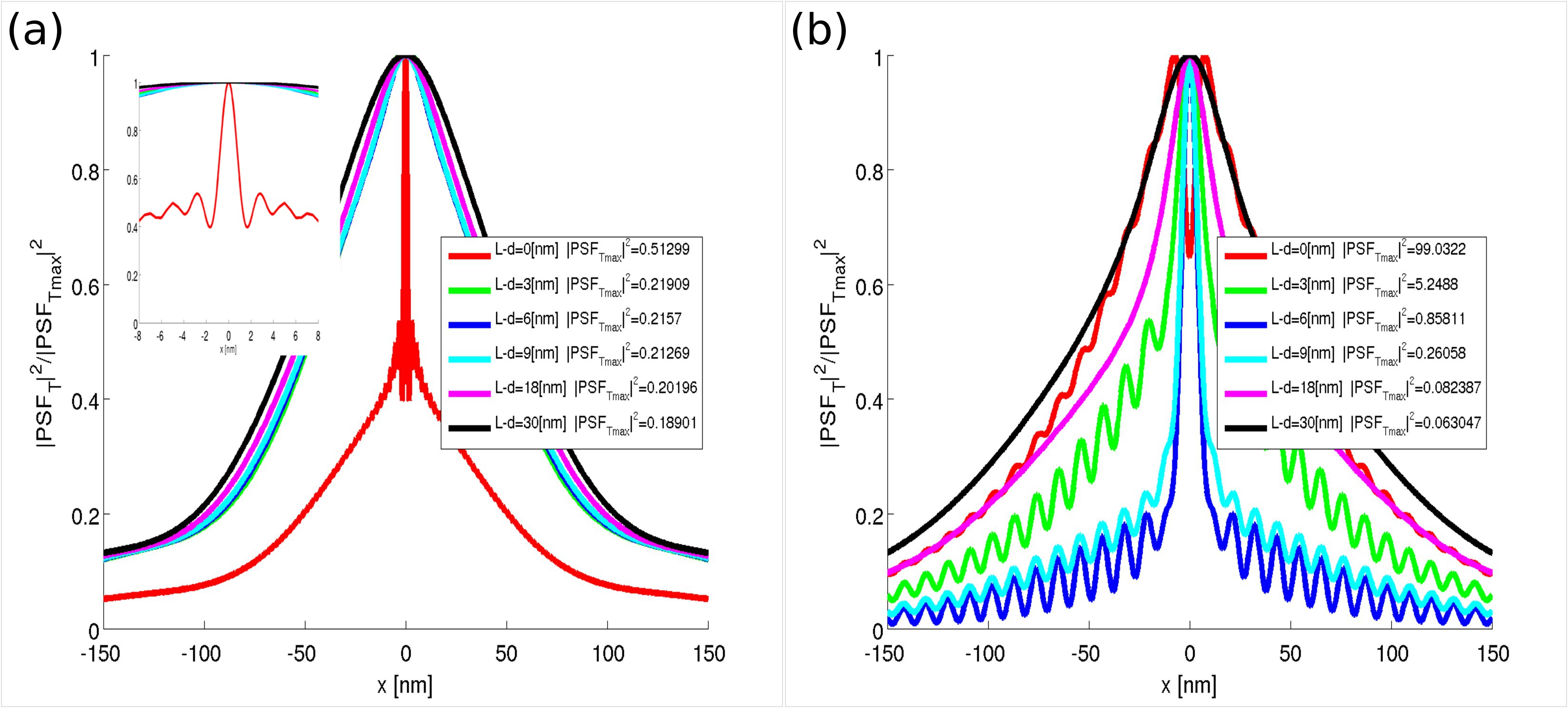}
\par\end{centering}

\caption{Normalized $PSF_{T}(x)$ calculated by Eq. \eqref{eq:PSFT}, as a
function of the transverse coordinate for several imaging distances.
The imaging distance $L-d$ and the normalization factor are shown
in the legend. The slab thickness is $d=5$ nm. (a) $\omega=1.217\omega_{p}$.
The inset shows a magnification of the PSF in the center. (b) $\omega=0.7071\omega_{p}$.
\label{fig:PSF_Ag_5}}
 
\end{figure}

The relatively large imaginary component of $k_{z,L}$, sets a practical
limitation on the slab thickness, preventing the use of slabs thicker
than $\sim10$ nm. Indeed, this thickness is known to be the critical
dimension for observing non-local effects in Ag \cite{yanai}. By
increasing $d$ beyond $3$ nm, the PSF becomes wider, and the transmission
magnitude reduces. This is shown in Fig. \ref{fig:PSF_Ag_5} presenting
the calculated $PSF_{T}$ for $d=5$ nm and two frequencies, $\omega=1.217\omega_{p}$
(corresponding to the FP resonance with $N=15$) and $\omega=0.7071\omega_{p}$.
The FWHM of $PSF_{T}$ is now $\sim6$ nm for the longitudinal FP
resonance case with $L-d=0$, which is about $3$ times larger than
the FWHM obtained for $d=3$ nm. For the plasmonic resonance case,
the increase of the layer thickness has a more moderate effect. For
the case shown in Fig. \ref{fig:PSF_Ag_3}(c), with $L-d=6$ nm, the
FWHM is $8$ nm, while for the case shown in Fig. \ref{fig:PSF_Ag_5}(b)
with $L-d=6$ nm, the FWHM is found increase only slightly, to $\sim10$
nm. 
\begin{figure}
\begin{centering}
\includegraphics[scale=0.08]{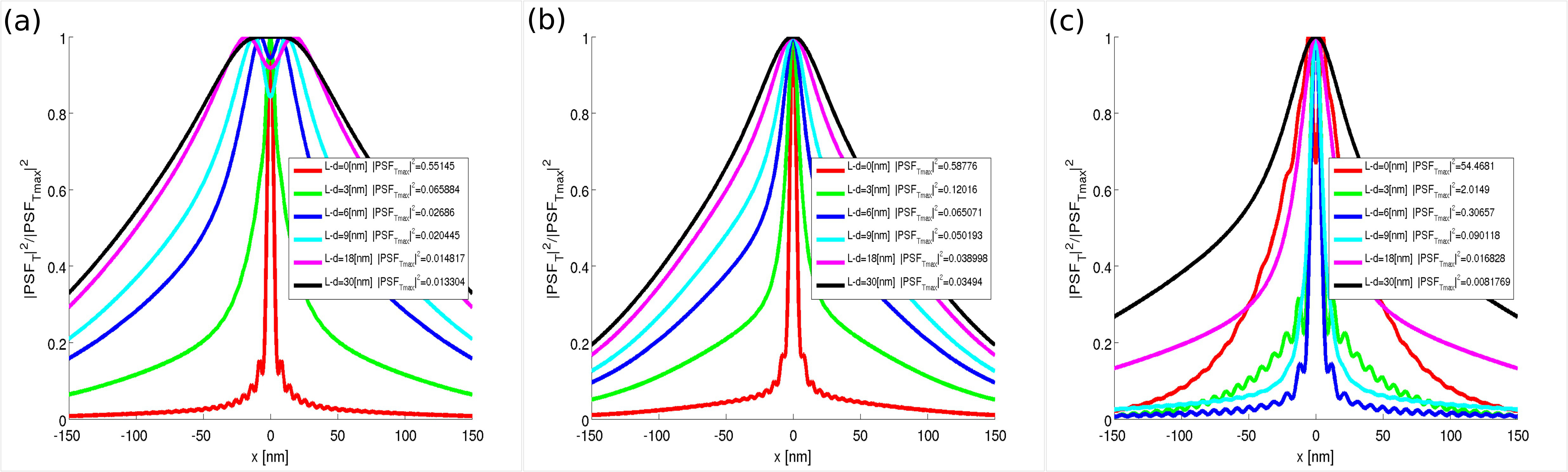}
\par\end{centering}

\caption{Normalized $PSF_{T}(x)$ calculated by Eq. \eqref{eq:PSFT}, as function
of the transverse coordinate for several imaging distances, calculated
for a K layer with $d=3$ nm. The imaging distance $L-d$ and the
normalization factor are shown in the legend. (a) $\omega=1.062\omega_{p}$
($N=3$) (b) $\omega=1.4915\omega_{p}$ ($N=9$) (c)$\omega=0.7071\omega_{p}$(plasmonic
resonance).\label{fig:PSF_K_3}}
 
\end{figure}

Finally, we consider the possibility of using a slab made of potassium
(K) rather than Ag. We assume the following parameters for K \cite{blaber}:
$\omega_{p}=3.72$ eV, $\gamma=(1/200)\omega_{p}$ , and $v_{F}=2.86\times10^{-3}c$.
Because the plasma frequency of K is lower than that of Ag and is
in the soft UV range, this metal might be more appealing for practical
demonstrations of the SDL imaging effect in the UV, e.g. for lithography
applications. In Fig. \ref{fig:PSF_K_3}(a,b) we show the PSF for
the $N=3$ ($\omega=1.062\omega_{p}$) and $N=9$ ($\omega=1.4915\omega_{p}$)
FP resonances, calculated for a K layer thickness of $d=3$ nm. In
Fig. \ref{fig:PSF_K_3}(c) we show the PSF for the plasmonic resonance
case ($\omega=0.7071\omega_{p}$). Because of the higher ohmic losses
of this metal, the PSF in all three cases is evidently wider as compared
to the Ag case. Yet, it may still be a more practical choice from
the technological point of view \cite{Anderegg}.

\section{Conclusion\label{sec:conclusions}}

In conclusion, we have analyzed the scenario of SDL imaging based
on longitudinal modes excited in a spatially dispersive metallic slab
above the metal plasma frequency. Our results show that in this regime,
for frequencies satisfying $k_{z,L}d/\pi=N$ ($N$ is an odd number)
a transmission spectrum nearly flat with $k_{x}$ is obtained, resulting
in an ultra narrow PSF for the case of full contact ($L-d=0$). This
PSF is significantly more narrow than the one obtained with the conventional
plasmonic based \textquotedblleft{}poor man\textquoteright{}s lens\textquotedblright{}
operated near $\omega_{p}/\sqrt{2}$. However, when $L-d$ is larger
than few nanometers, the situation is inverted and the plasmonic \textquotedblleft{}poor
men lens\textquotedblright{} becomes a much better choice as it provides
much narrower PSF. Another interesting difference is related to the
wavelength of choice. With our proposed approach, the frequency of
operation can be tuned (either by changing the layer thickness, or
by choosing a higher order FP resonance), while for the SP resonance
case the operation frequency is fixed in the vicinity of $\sim\omega_{p}/\sqrt{2}$.
We believe that the proposed concept could have several useful applications
for fields such as nano-lithography, sensing, and microscopy, to name
a few.\appendix

\section{Calculation of the transmission amplitude for a spatially dispersive
slab}

For completeness, we derive here the expression for the transmission
amplitude for the case of a spatially dispersive slab surrounded by
local media. Similar derivations are available for the slab case \cite{melnyk,forstmann-book,david_sci_rep},
and the more general periodic case \cite{yanai}. The slab geometry
and the field amplitudes are depicted in Fig. \ref{fig:appendix-schematic}.
The incident and reflected amplitudes are denoted by $E_{0}$ and
$E_{r}$ respectively. In the slab, we define two transverse mode
amplitudes $E_{1}$ and $E_{3}$, propagating and counter-propagating
respectively, and two longitudinal mode amplitudes $E_{2}$ and $E_{4}$.
The transmitted mode amplitude is $E_{t}$. To derive the complex
transmission amplitude $T=E_{t}/E_{0}$, we match all fields according
to the assumed ABC. For this geometry, there are three types of incidence
of the various modes:
\begin{enumerate}
\item Transverse mode incident from a spatially non-dispersive medium onto
a spatially dispersive medium
\item Transverse mode incident from a spatially dispersive medium onto a
spatially non-dispersive medium
\item Longitudinal mode incident from a spatially dispersive medium onto
a spatially non-dispersive medium
\begin{figure}[H]
\begin{centering}
\includegraphics[scale=0.5]{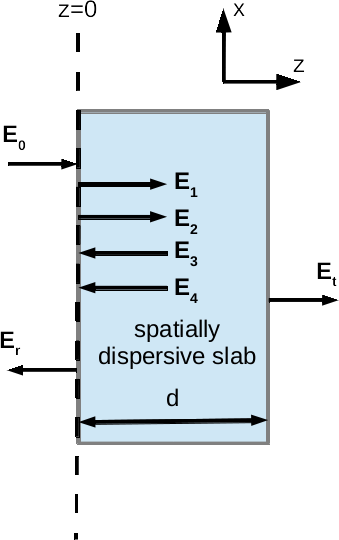}
\par\end{centering}

\caption{Schematic showing the field components of the various modes for the
scenario of a spatially dispersive slab surrounded by local media.
\label{fig:appendix-schematic} }
\end{figure}

\end{enumerate}
Employing the ABC, the transmission and reflection coefficients for
these three incidence types can be derived: \begin{subequations}\label{eq:trans_refl_coef}

\begin{equation}
R_{t1}=\frac{\varepsilon_{T}k_{z0}-k_{z,T}+(\varepsilon_{T}-1)k_{x}^{2}/k_{z,L}}{\varepsilon_{T}k_{z0}+k_{z,T}-(\varepsilon_{T}-1)k_{x}^{2}/k_{z,L}}
\end{equation}

\begin{equation}
T_{t1}=\frac{2k_{z0}}{\varepsilon_{T}k_{z0}+k_{z,T}-(\varepsilon_{T}-1)k_{x}^{2}/k_{z,L}}
\end{equation}

\begin{equation}
T_{l1}=\frac{2k_{z0}(\varepsilon_{T}-1)}{\varepsilon_{T}k_{z0}+k_{z,T}-(\varepsilon_{T}-1)k_{x}^{2}/k_{z,L}}
\end{equation}

\begin{equation}
R_{l2}=\frac{2k_{z,T}(\varepsilon_{T}-1)}{\varepsilon_{T}k_{z0}+k_{z,T}-(\varepsilon_{T}-1)k_{x}^{2}/k_{z,L}}
\end{equation}

\begin{equation}
R_{t2}=\frac{k_{z,T}+(\varepsilon_{T}-1)k_{x}^{2}/k_{z,L}-\varepsilon_{T}k_{z0}}{\varepsilon_{T}k_{z0}+k_{z,T}-(\varepsilon_{T}-1)k_{x}^{2}/k_{z,L}}
\end{equation}

\begin{equation}
T_{t2}=\frac{2k_{z,T}\varepsilon_{T}}{\varepsilon_{T}k_{z0}+k_{z,T}-(\varepsilon_{T}-1)k_{x}^{2}/k_{z,L}}
\end{equation}

\begin{equation}
R_{l3}=-\frac{k_{z,T}+(\varepsilon_{T}-1)k_{x}^{2}/k_{z,L}+\varepsilon_{T}k_{z0}}{\varepsilon_{T}k_{z0}+k_{z,T}-(\varepsilon_{T}-1)k_{x}^{2}/k_{z,L}}
\end{equation}

\begin{equation}
R_{t3}=-\frac{2k_{x}^{2}/k_{z,L}}{\varepsilon_{T}k_{z0}+k_{z,T}-(\varepsilon_{T}-1)k_{x}^{2}/k_{z,L}}
\end{equation}

\begin{equation}
T_{t3}=-\frac{2\varepsilon_{T}k_{x}^{2}/k_{z,L}}{\varepsilon_{T}k_{z0}+k_{z,T}-(\varepsilon_{T}-1)k_{x}^{2}/k_{z,L}}
\end{equation}
\end{subequations}For these nine coefficients, $R$ and $T$ are
for reflection and transmission respectively. The subscripts $l$
and $t$ are for longitudinal and transverse modes respectively, and
the subscripts $1,2$ and $3$ correspond to each of the three incidence
types respectively. For example, $R_{t3}$ stands for the reflection
coefficient of a transverse mode due to a longitudinal mode incident
from a spatially dispersive medium onto air. We define six additional
auxiliary transmission and reflection coefficients, based on those
defined in Eq.\eqref{eq:trans_refl_coef}: $\tilde{T}_{t2}=T_{t2}\exp(j\Phi_{T})$,
$\tilde{T}_{t3}=T_{t3}\exp(j\Phi_{L})$, $\tilde{R}_{t2}=R_{t2}\exp(j2\Phi_{T})$,
$\tilde{R}_{t3}=R_{t3}\exp(j[\Phi_{T}+\Phi_{L}])$, $\tilde{R}_{l2}=R_{l2}\exp(j[\Phi_{T}+\Phi_{L}])$
and $\tilde{R}_{l3}=R_{l3}\exp(j2\Phi_{L})$. Here, $\Phi_{T}=k_{z,T}d$
and $\Phi_{L}=k_{z,L}d$ are the phases accumulated by the transverse
and longitudinal modes respectively when transversing the slab. With
these definitions, we write a system of equation describing the mode
amplitudes:

\begin{equation}
\left[\begin{array}{ccccccc}
0 & 0 & 0 & T_{t2} & T_{t3} & 0 & R_{t1}\\
0 & 0 & 0 & R_{t2} & R_{t3} & 0 & T_{t1}\\
0 & 0 & 0 & R_{l2} & R_{l3} & 0 & T_{l1}\\
0 & \tilde{R}_{t2} & \tilde{R}_{t3} & 0 & 0 & 0 & 0\\
0 & \tilde{R}_{l2} & \tilde{R}_{l3} & 0 & 0 & 0 & 0\\
0 & \tilde{T}_{t2} & \tilde{T}_{t3} & 0 & 0 & 0 & 0\\
0 & 0 & 0 & 0 & 0 & 0 & 1
\end{array}\right]\left[\begin{array}{c}
E_{r}\\
E_{1}\\
E_{2}\\
E_{3}\\
E_{4}\\
E_{t}\\
E_{0}
\end{array}\right]=\left[\begin{array}{c}
E_{r}\\
E_{1}\\
E_{2}\\
E_{3}\\
E_{4}\\
E_{t}\\
E_{0}
\end{array}\right]\label{eq:system_of_equations}
\end{equation}
The solution of the mode amplitudes are obtained by solving Eq. \eqref{eq:system_of_equations}
\begin{subequations}\label{eq:mode-amplitudes}
\begin{equation}
E_{1}=\left[\frac{T_{t1}\beta+T_{l1}\gamma}{\alpha\beta-\gamma\delta}\right]E_{0}
\end{equation}

\begin{equation}
E_{2}=\left[\frac{T_{l1}\alpha+T_{t1}\delta}{\alpha\beta-\gamma\delta}\right]E_{0}
\end{equation}

\begin{equation}
E_{3}=\tilde{R}_{t2}\left[\frac{T_{t1}\beta+T_{l1}\gamma}{\alpha\beta-\gamma\delta}\right]E_{0}+\tilde{R}_{t3}\left[\frac{T_{l1}\alpha+T_{t1}\delta}{\alpha\beta-\gamma\delta}\right]E_{0}
\end{equation}
\begin{equation}
E_{4}=\tilde{R}_{l2}\left[\frac{T_{t1}\beta+T_{l1}\gamma}{\alpha\beta-\gamma\delta}\right]E_{0}+\tilde{R}_{l3}\left[\frac{T_{l1}\alpha+T_{t1}\delta}{\alpha\beta-\gamma\delta}\right]E_{0}
\end{equation}
\begin{equation}
E_{t}=\tilde{T}_{t2}\left[\frac{T_{t1}\beta+T_{l1}\gamma}{\alpha\beta-\gamma\delta}\right]E_{0}+\tilde{T}_{t3}\left[\frac{T_{l1}\alpha+T_{t1}\delta}{\alpha\beta-\gamma\delta}\right]E_{0}
\end{equation}

\begin{equation}
E_{t}=\left[\frac{\tilde{T}_{t2}T_{t1}\beta+\tilde{T}_{t2}T_{l1}\gamma+\tilde{T}_{t3}T_{l1}\alpha+\tilde{T}_{t3}T_{t1}\delta}{\alpha\beta-\gamma\delta}\right]E_{0}
\end{equation}
\end{subequations}Here we define: $\alpha=\left(1-R_{t2}\tilde{R}_{t2}-R_{t3}\tilde{R}_{l2}\right)$,
$\beta=\left(1-R_{l2}\tilde{R}_{t3}-R_{l3}\tilde{R}_{l3}\right)$,$\gamma=\left(R_{t2}\tilde{R}_{t3}+R_{t3}\tilde{R}_{l3}\right)$
and $\delta=\left(R_{l2}\tilde{R}_{t2}+R_{l3}\tilde{R}_{l2}\right)$.
Eq. \eqref{eq:mode-amplitudes} describes all field amplitudes for
this geometry.
\begin{acknowledgments}
This research was supported by the AFOSR.\end{acknowledgments}

\end{document}